\documentstyle[12pt]{article}
\newcommand{\bee}{\begin{equation}}
\newcommand{\ee}{\end{equation}}
\newcommand{\ba}{\begin{array}}
\newcommand{\ea}{\end{array}}
\newcommand{\bea}{\begin{eqnarray}}
\newcommand{\eea}{\end{eqnarray}}

\begin{document}
\thispagestyle{empty}
\begin{flushright}
MPI-PhT/94-07\\
AZPH-TH/94-03\\
February 1994
\end{flushright}
\bigskip\bigskip\begin{center}
{\bf \Huge{Super-Instantons in Gauge Theories and}}\vskip2mm
{\bf \Huge{Troubles with Perturbation Theory}}
\end{center}
\vskip 1.0truecm
\centerline{\bf
Adrian Patrascioiu\footnote{Permanent address: Physics Department,
University of Arizona, Tucson, AZ 85721, \hspace*{6mm}U.S.A.}}
\vskip5mm
\centerline{and}
\vskip5mm
\centerline{\bf Erhard Seiler}
\vskip5mm
\centerline{Max-Planck-Institut f\"ur
 Physik}
\centerline{ -- Werner-Heisenberg-Institut -- }
\centerline{F\"ohringer Ring 6, 80805 Munich, Germany}
\vskip 2cm
\bigskip \nopagebreak \begin{abstract}
\noindent
In gauge theories with continuous groups there exist classical
solutions whose energy vanishes in the thermodynamic limit (in any
dimension). The existence of these {\it super-instantons} is intimately
related to the fact that even at short distances perturbation theory
can fail to produce unique results. This problem arises only in non-Abelian
models and only starting at $O(1/\beta^2)$.
\end{abstract}
\vskip 2cm

\newpage\setcounter{page}1


The standard belief in Quantum Field Theory
is that in asymptotically free theories
perturbation theory predicts correctly the short distance behavior of
the Green's functions -- and that the hard part is understanding their
infrared behavior. In this letter we will show via some simple
calculations that this belief is erroneous:
in these theories the ultraviolet (UV) and infrared
(IR) effects are entangled in such a way that in fact perturbation
theory fails even at short distances; in particular the main assumption
underlying the applications of the operator product expansion
(regarding a certain factorization of UV and IR effects) is invalid in
theories such as QCD.

Since our claim runs against the established beliefs, we would like to
provide a clear exposition of the issues involved. Consequently, at the
risk of boring some readers, we will repeat the general discussion needed
to comprehend what is the trouble \cite{FF}.
We begin by reminding the reader
why one needs a non-perturbative definition of a quantum field theory.
Firstly, if one claims, as is usually done, that perturbative QCD
(PQCD) is correct at short
distances, what exactly is one claiming? The temptation is usually to
appeal to experiment, but that is silly since most gauge
models have nothing
to do with nature and one could still ask the question. The only way to
make sense of such a claim is to give the theory a non-perturbative
definition and to argue that in a certain regime perturbation theory
produces a good approximation.

A second reason underlying the need for a non-perturbative definition
of a quantum field theory comes from the fact that correct or not,
perturbation theory provides us with answers in the form of divergent
(non-convergent) series. Whereas a unique numerical answer can be associated
with any convergent series, divergent power series have no intrinsic meaning,
but become meaningful only if some external (non-perturbative) definition is
provided. Currently popular attempts to `sum up' PT are therefore
meaningless in the absence of a nonperturbative definition of the theory.

Therefore both to understand the meaning of the question `is PQCD
correct at short distances?' and to make sense of the predictions of
PQCD, one needs a non-perturbative definition of QCD. We will adopt
the lattice (LQCD) approach. Some readers may conclude that our
unpleasant conclusions are a consequence of our use of LQCD. Although
that is logically possible, unless an alternative non-perturbative
framework is proposed and the troubles appearing in LQCD are shown to
disappear, such an excuse cannot be taken seriously.

Following Wilson \cite{W}, we consider a regular cubic lattice in
$D$ dimensions and define the partition function as

\bee
    Z_\Lambda=\int\prod_{x,\mu}dU_\mu(x)e^{{\beta\over N}
\sum_{plaq} {\rm tr}\  U_{plaq}}
\ee
Here $U\in SU(N)$ and
$\Lambda\subset {\bf Z}^D$ is a hypercube of linear extension $L$.
To fully specify the problem we must impose some boundary conditions
(b.c.). We will choose the following b.c.: in the hyperplane $x_1=0$, all
$U_{plaq}=1$, while in the hyperplane $x_1=L$ all $U_{link}$ are free
variables;
in the remaining directions $\Lambda$ is a hypertorus (periodic b.c.).

We chose these b.c. because they are easily compatible with the maximally
axial gauge. For simplicity we will discuss the case $D=3$, the extension
to higher $D$ being obvious. For our b.c. the maximally axial gauge amounts
to $U_1(x)=1$, $x\in\Lambda$ and $U_2(0,i,j)=1=U_3(0,i,j)$ for all $(i,j)$.
Fixing the gauge is
necessary if we wish to do perturbation theory (PT). Indeed as
$\beta\to\infty$, once the gauge has been fixed and the b.c. specified,
the system will perform small oscillations around one or several
(classical) configurations. PT is simply the saddle point
approximation built
around these classical configurations. For $L$ finite PT is correct in the
following sense: let $G(\beta)$ be the
expectation value of some gauge invariant
observable. Then for $\beta\to\infty$ PT produces the correct asymptotic
expansion of $G(\beta)$ in powers of $1/\beta$. The mathematical meaning of the
statement is that if one truncates the PT formal series at order $k$, the
error is $o({1/\beta^k})$.

Therefore to verify the correctness of a certain PT prediction one needs
not only a computation of the coefficients entering the expansion of
$G(\beta)$, but also a bound on the truncation error. That is easy to do for
finite $L$. Unfortunately for
non-Abelian groups the best estimates of the truncation error are such
that they diverge as $L\to\infty$ (for Abelian groups a bound uniform in
$L$ has been shown to exist \cite{Bric}). Thus it is mathematically
unknown if the PT expansion of a short distance quantity, such as the
plaquette energy, remains valid as $L\to\infty$.

Since many things which are true are nevertheless hard to prove, one
may wonder if there is any cause for doubting PT. As emphasized several
years ago by Patrascioiu \cite{Pat}, there is good reason to do so.
The saddle point approximation of an integral amounts to replacing the
integrand by a Gaussian plus corrections.
Intuitively one would guess that such an
approximation
is good provided the integrand is sharply peaked and the peak is
sufficiently far
from the boundaries of the integration region. Now in the maximally
axial gauge, Patrascioiu showed that as $L\to\infty$, the integrand
becomes arbitrarily flat; that is in that gauge, for any $D$, no matter
how large $\beta$ is, for $L$ sufficiently large one will encounter large
fluctuations. This statement is much stronger than Elitzur's theorem
\cite{Eli}:
it says that in gauge theories, for any $D$, in the maximally axial
gauge there is no long range order for any finite $\beta$. Since PT is
an expansion in small deviations from an ordered state and such a
state does not exist on an infinite lattice, Patrascioiu concluded
that even at short distances PT was highly suspicious.

One may wonder if Patrascioiu's conclusion could not be dismissed as a
gauge artifact, since it is well known that at least for $D>2$, the
IR divergences which PT displays in this gauge disappear in other gauges.
Unfortunately the answer is that this is not a gauge choice artifact, but
a genuine difficulty of gauge theories. To see that consider firstly the
fact that the asymptotic expansion of $G(\beta)$, if it exists, it is
unique. Now for finite $L$, the maximally axial gauge is definitely usable.
Whatever answers PT produces in this gauge they must represent the true
asymptotic expansion of $G(\beta)$ and any other gauge choice is bound to
reproduce them. Since the PT answers in any other gauge must agree with
the PT answers in the maximally axial gauge for finite $L$, they must also
agree for $L\to\infty$. Therefore if PT is not uniformly valid for $L\to
\infty$, that fact has nothing to do with the use of the maximally axial
gauge.

A direct corroboration of this conclusion comes from the fact
that while trying to avoid the large fluctuations problem of the maximally
axial gauge by using the Landau gauge, Zwanziger \cite{20}
discovered a new trouble
which makes PT suspicious: as $L\to\infty$, the boundary of the integration
region (the `fundamental modular domain')
collides with the position of the peak of the integrand.

To recap the discussion presented so far, there are good reasons to
suspect that in gauge theories, for any $D$, taking the termwise $L\to
\infty$ limit of the PT coefficients may actually produce
incorrect answers in non-Abelian models. To go beyond this point we
would need an actual estimate of the truncation error, a task beyond
our present abilities. Following a similar ruse we used recently in
the $2D$ non-linear $\sigma$-models \cite{Super}, we will avoid that task
by appealing to a special property which these models have and which
a correct PT computation should also exhibit. That property is the
absence of symmetry breaking, which in the $2D$ $\sigma$-models is
the Mermin-Wagner theorem and in gauge theories in the maximally
axial gauge follows from Patrascioiu's observations \cite{Pat}. This property
implies that in the infinite volume limit, in the maximally axial gauge
the expectation value of the energy of a plaquette
located in the middle of the lattice is the same
whether or not we fix an additional
link variable $U_2(m)$ at some random value. The correctness of this
statement is easy to verify via the convergent strong coupling
cluster expansion \cite{OS}.
For illustrative purposes we chose to verify it also numerically and
present the results in Tab.1 for $SU(2)$ at $\beta=3.0$. The three plaquettes
investigated are shown in Fig.1 and we chose $U_2(m)=1$. The data show
clearly that as $L$ increases the three expectation values converge to the
same value, representing the thermodynamic value of this observable.

Next we would like to inquire whether the PT answers respect this
property of the model. For that purpose we must compare the limit
$L\to\infty$ in two PT computations: the first one with what we shall call
the Dirichlet b.c. described above (below eq.(1)) and the second one
with what we shall call super-instanton (s.i.) b.c., namely Dirichlet
plus $U_2(m)=1$. For the Dirichlet case the algebra has been carried out
to a large extent by M\"uller and R\"uhl \cite{MR}, whose procedure we
followed. It yields the following infinite volume expression for the
PT expansion of the plaquette energy (for $SU(2)$ in $D=3$)

\bee
 \langle E\rangle\equiv
\langle {\rm tr}\ U_{plaq}\rangle =1-c_1/\beta-c_2/\beta^2+...
\ee
\bee
         c_1=1,\ \ \  c_2=.23
\ee
We have compared the value of $c_2$ obtained by us in this maximally axial
gauge with that obtained by Wohlert et al \cite{WW} in covariant
gauges and they agree.

To obtain the PT coefficients with the s.i. b.c. we need a modified
propagator, which vanishes on the link $(2,m)$. Following a suggestion
made to us by Sokal (private communication), given a certain propagator
$G(x,y)$ the combination

\bee
   \tilde G(x,y)=G(x,y)-G(x,0)G(0,y)/G(0,0)
\ee
will be the propagator with the additional b.c. that it vanishes at $0$.
Therefore out of the previous
(Dirichlet) propagators one can easily construct the s.i.~propagators
and thus compute the PT coefficients for these b.c.. We computed the
the coefficients $c_1$ and $c_2$ and found that $\lim_{L\to\infty} c_1(L)=1$
even with s.i. b.c..
The results of the computation of $c_1(l)$ and $c_2(L)$ with s.i. b.c.
for various $L$ values are shown in Tab.2. The remarkable
finding is that the two plaquettes $P_1$ and $P_2$ (see Fig.1),
sharing the same frozen link $U_2(m)=1$ have PT coefficients $c_2$
converging to different values -- and in fact only the limit of
the coefficient of $P_2$ agrees with that obtained with Dirichlet b.c..
The data for $P_1$ are perfectly described by

\bee
c_2(L)=.40633-1.16026/L-.49819/L^2
\ee

As the discussion of the vacuum structure below shows, the mechanism
responsible for this effect operates in any dimension $D\ge2$.
Therefore, as stated in the introduction, in non-Abelian models PT
fails to reproduce the true properties of the model, such as the
independence of the expectation value of the energy upon the b.c. used
to reach the thermodynamic limit. This effect occurs only at $O(1/\beta^2)$
because only from that order on PT computations involve loop-integrations
which mix low and high momenta. Also the effect does not occur in
Abelian theories, which contain no cancelling IR divergences (the action
depends only on gradients and the link measure is flat). Finally, let
us notice that a similar effect should appear in a plaquette-plaquette
2-point function from which one could determine
the Callan-Symanzik $\beta$-function. Thus the PT computation of the latter
probably suffers from the same ambiguities and is not universal, as usually
claimed. We verified that this actually happens in the $2D$ nonlinear
$\sigma$-model \cite{Super}.

A few years ago Gribov \cite{Grib} suggested that the long distance
behavior of ${\rm QCD}_4$ is different from the usual picture of
`infrared slavery', but that nevertheless PT describes correctly
its short distance behavior. Our present results show however that even
at short distances PT should not be trusted.
On the other hand, what we found here
lends support to the scenario presented by us \cite{Exp}, according
to which ${\rm LQCD}_4$ undergoes a zero temperature deconfining
transition at finite $\beta$; such a transition would lead
to a slower variation of $\alpha_s$ with the energy than
predicted by PQCD, a prediction which has since found some experimental
support from the LEP data.

Some readers may wonder if there is a connection between the troubles
with PT we have been pointing out and the claims \cite{LP} that
$\beta$ is a `bad expansion parameter' and needs to be replaced by an
`improved one'.
The answer is no: the question
we are addressing is not whether $\beta$ is a good
expansion parameter, but rather whether conventional PT gives the correct
asymptotic expansion.

Also some readers may wonder if our results are not contradicted by
the constructions of continuum Yang-Mills theory by Magnen et al
\cite{Poly}, which also claim to establish asymptotic freedom.
These constructions work in a small volume, precisely to avoid
the large infrared fluctuations which are responsible for the effects
we are describing in this paper.

Next let us explain the connection between these troubles of PT and the
structure of the vacuum. In the maximally axial gauge the trouble
arises because as $L$ grows the system becomes less and less ordered. How
does that happen? To understand that let us consider the energy of the
classical configuration obtained by using Dirichlet b.c. (in the
maximally axial gauge) and fixing $U_2(L/2,0,0)=V$
for some $V=\exp(i\vec{\tau\over 2}\cdot\vec C)\in SU(2)$. We cannot
write the analytic expression for this classical configuration, but its
general features are easy to comprehend. To that end let us write

\bee
   U_\mu(x)=\exp(i {\vec\tau\over 2}\cdot \vec A_\mu(x))
\ee
The b.c. require $\vec A_2(0,0,0)=0$ and
$\vec A_2(L/2,0,0)=\vec C$. Consistent with these
b.c., suppose that

\bee
            \vec A_2(x_1,0,0)={x_1\over L}\vec C
\ee
The total energy of the (1,2)-plaquettes lying between $x_1=0$ and $x_1=L/2$
and having $x_2=0=x_3$ is $O(C^2/L)$, hence it vanishes as $L\to\infty$. Of
course
other plaquettes will also carry energy, but the total energy of the
configuration will nevertheless vanish as $1/L$ for any $D$ (we
verified this numerically for $D=3$). Indeed a gauge
invariant description of the configuration we are discussing is this:
there is a thin Wilson loop of length $L/2$ (width 1 lattice spacing)
having the value $\exp(i\vec{\tau\over 2}\cdot \vec C)$.
The magnetic field is $O(|\vec C|/L)$ and dying off as $r^{-D}$
as one goes transversely away from the loop.
Consequently even though in this configuration $A_\mu=O(1)$,
its energy vanishes as $L\to\infty$.

This is why we baptised these classical configurations super-instantons.
Since they have arbitrarily low energy (and a lot of entropy) they will occur
copiously in any gauge theory at weak coupling. In fact the typical
configuration at weak coupling could be regarded as a gas or liquid of
super-instantons. This picture differs from the so-called `spaghetti vacuum'
\cite{NO}, which has higher free energy.
In the $2D$ $O(N)$ $\sigma$-models certain percolation results \cite{Super}
allow one to conclude that if the typical configuration looks
like a gas of super-instantons, then the model must be massless. In
gauge theories such a connection is missing so far.

Nevertheless, since the super-instantons are practically degenerate
with the configuration $\vec A=0$, and since they are classical solutions,
one must consider fluctuations around them. Our calculations correspond to
PT around a super-instanton with $\vec C=0$. In general one should
expect the answer to be dependent on $|\vec C|$, since
in the maximally axial gauge the IR divergencies are $O(L)$,
whereas the new vertices induced by the super-instanton field are
$O(1/L)$.

A.P.~is grateful to the Alexander-von-Humboldt Foundation for a
Senior U.S. Scientist Award and to W.Zimmermann for his hospitality at
the Max-Planck-Institute. We thank P.Weisz for information regarding
PT in covariant gauges.

\pagebreak
\vskip1cm
\noindent
{\bf Fig.1}: The lattice in $3D$. On the links drawn in heavy lines
$U_{link}=1$. On the link common to the plaquettes $P_1$, $P_2$ and $P_3$,
$U_{link}=1$ for s.i. b.c..


\noindent
{\bf Tab.1:} Plaquette energies in $3D$ for the plaquettes $P_1$, $P_2$
and $P_3$ (see fig.1) (Monte Carlo data at $\beta=3.0$).

  \medskip
  \vbox{\offinterlineskip\halign{
  \strut\vrule#&\quad $#$\quad&\vrule\hskip1pt\vrule#&&\quad $#$\hskip5pt
 &\vrule#\cr
  \noalign{\hrule}
  &L&& 4&&  6&&  8&&  12&& 20&& 30&\cr
  \noalign{\hrule}
  &P_1&&.809(25)&&.738(30)&&.689(39)&&.657(50)&&.621(39)&&.632(66)&\cr
  \noalign{\hrule}
  &P_2&&.638(72)&&.614(79)&&.615(81)&&.617(73)&&.626(28)&&.616(45)&\cr
  \noalign{\hrule}
  &P_3&&.647(61)&&.632(49)&&.631(36)&&.626(32)&&.624(14)&&.629(19)&\cr
  \noalign{\hrule}}}

\vskip2cm
\noindent
{\bf Tab.2a:} The PT coefficients $c_1(L)$ for the energies of the plaquettes
$P_1$ and $P_2$ computed with super-instanton b.c..

  \medskip
  \vbox{\offinterlineskip\halign{
  \strut\vrule#&\quad $#$\quad&\vrule\hskip1pt\vrule#&&\quad $#$\hskip5pt
  &\vrule#\cr
   \noalign{\hrule}
   &L&& 8&& 10&& 12&& 16&& 20&& 30&\cr
   \noalign{\hrule\vskip1pt\hrule}
   &P_1&&.7587&&.8051&&.8364&&.8762&&.9004&&.9331&\cr
   \noalign{\hrule}
   &P_2&&.9967&&.9968&&.9971&&.9978&&.9981&&.9987&\cr
   \noalign{\hrule}}}

\vskip2cm
\noindent
{\bf Tab.2b:} The PT coefficients $c_2(L)$ for the energies of the
plaquettes $P_1$ and $P_2$ computed with super-instanton b.c..

  \medskip
  \vbox{\offinterlineskip\halign{
  \strut\vrule#&\quad $#$\quad&\vrule\hskip1pt\vrule#&&\quad $#$\hskip5pt
  &\vrule#\cr
   \noalign{\hrule}
   &L&& 8&& 10&& 12&& 16&& 20&& 30&\cr
   \noalign{\hrule\vskip1pt\hrule}
   &P_1&&.2536&&.2852&&.3061&&.3320&&.3472&&.3669&\cr
   \noalign{\hrule}
   &P_2&&.1594&&.1893&&.2050&&.2201&&.2268&& &\cr
   \noalign{\hrule}}}

\end{document}